\documentclass[useAMS,usenatbib,usegraphicx]{mn2e}

\usepackage{lscape}
\usepackage{color}
\usepackage{subfigure}
\usepackage{amssymb}
\usepackage{amsmath}
\usepackage{url}
\usepackage{amsfonts}
\usepackage{amsbsy}
\usepackage{graphicx}
\usepackage{subfigure}
\usepackage{verbatim}
\usepackage{multicol}

\title[Torus Constrains of Compton-Thick Seyfert 2 Mrk\,3]{High Spatial Resolution of the Mid-Infrared
Emission of Compton-Thick Seyfert 2 Galaxy Mrk\,3}
\author[Dinalva A. Sales et al.]{Dinalva A. Sales$^{1,2}$\thanks{dinalvaires@gmail.com};D. Ruschel-Dutra$^{2}$; M. G. Pastoriza$^{2,3}$
; R. Riffel$^{2}$; Cl\'audia Winge$^{4}$\\
$^{1}$Department of Physics, Rochester Institute of Technology, 84 Lomb Memorial Drive, Rochester, NY 14623, USA\\
$^{2}$Departamento de Astronomia, Universidade Federal do Rio Grande do Sul. 9500 Bento Gon\c calves, Porto Alegre, 91501-970, Brazil\\
$^{3}$Conselho Nacional de Desenvolvimento Cient\' ifico e Tecnol\' ogico, Brasilia, 71605-001, Brazil\\
$^{4}$Gemini Observatory, c/o Aura, Inc., Casilla 603, La Serena, Chile}

\begin{document}

\date{Accepted 1988 December 15. Received 1988 December 14; in original form 1988 October 11}

\pagerange{\pageref{firstpage}--\pageref{lastpage}} \pubyear{2012}

\maketitle

\label{firstpage}

\begin{abstract}

Mid-infrared (MIR) spectra observed with Gemini/Michelle were used to study the nuclear 
region of the Compton-thick Seyfert~2 (Sy~2) galaxy Mrk~3 at a spatial resolution of $\sim$200\,pc.
No polycyclic aromatic hydrocarbons (PAHs) emission bands were detected in the N-band spectrum 
of Mrk~3. However, intense [Ar{\sc\,iii]}\,8.99\,$\mu$m, [S{\sc\,iv]}\,10.5\,$\mu$m and 
[Ne{\sc\,ii]}\,12.8\,$\mu$m ionic emission-lines, as well as silicate absorption feature 
at 9.7$\mu$m have been found in the nuclear extraction ($\sim$200\,pc). We also present subarcsecond-resolution 
Michelle N-band image of Mrk\,3 which resolves its circumnuclear region. This diffuse MIR 
emission shows up as a wings towards East-West direction closely aligned with the S-shaped of the 
Narrow Line Region (NLR) observed at optical [O{\sc\,iii]}\,$\lambda$5007\AA\,\,image with 
\emph{Hubble/FOC}. The nuclear continuum spectrum can be well represented by a theoretical torus 
spectral energy distribution (SED), suggesting that the nucleus of Mrk~3 may host a dusty toroidal 
structure predicted by the unified model of active galactic nucleus (AGN). In addition, the 
hydrogen column density (N$_H\,=\,4.8^{+3.3}_{-3.1}\times\,10^{23}$\,cm$^{-2}$) estimated with a 
torus model for Mrk~3 is consistent with the value derived from X-ray spectroscopy. The torus model geometry of 
Mrk~3 is similar to that of NGC~3281, both Compton-thick galaxies, confirmed through fitting 
the 9.7$\mu$m silicate band profile. This results might provide further evidence that the 
silicate-rich dust can be associated with the AGN torus and may also be responsible for 
the absorption observed at X-ray wavelengths in those galaxies.

%After removal of the underlying galaxy contribution, the nuclear spectrum can be represented 
%by a Nenkova's {\sc clumpy} torus model, suggesting that the nucleus of Mrk~3 hosts a dusty toroidal structure with an angular cloud distribution of $\sigma = 15^{\circ}$, 
%observer's view angle of $i = 90^{\circ}$, and an outer radius of R$_{0}\sim$34\,pc. The derived column density along the observer line of sight is N$_H$\,=\,1.4\,$\times\,10^{24}$\,cm$^{-2}$ 
%and the MgSiO3 amorphous silicate density is 7.1\,$\times\,10^{-21}$\,Kg~m$^{-3}$. The torus properties of Mrk~3 are compared with those of NGC~3281. Our results
%indicate very different torus geometry for the both Mrk~3 and NGC~3281 galaxies, however, they provide further evidence that the silicate-rich dust is associated with the AGN torus and could also be responsible for the absorption observed at X-ray wavelengths, which classified these galaxies as
%Compton-thick sources.

\end{abstract}


\section{Introduction}

%%%%%%%%%%% MUDEI %%%%%%%%%%%%%%%%%%%%%%%%%%
The currently favoured unified models of active galactic nucleus (AGN) are ``orientation-based 
models''. They propose that the differences between different classes of objects arise 
because of their different orientations to the observer. These models propose the existence 
of a dense concentration of absorbing material in their central engine in a toroidal 
distribution, which blocks the broad line region (BLR) from the line of sight in Type~2 
objects \citep[see][for a review ]{antonucci93,urry95}. However, a not well understood, but 
key issue in AGN physics is the composition and nature of this dusty torus. 

For example, several of these models predict that the silicate emission/absorption 
features at 9.7$\mu$m and 18$\mu$m are related to the observers viewing angle, in the framework 
of the AGN unified model
\citep[e.g.][and references therein]{pier92,granato94,granato97,rowan95,nenkova02,nenkova08a,nenkova08b,schartmann05,schartmann08,dullemond05,fritz06,honig06,honig10,stalevski12,heymann12,efstathiou13}.
The strengths of these features are sensitive to the dust distribution and 
could be a direct evidence of a connection between mid-infrared (MIR) optically-thick galaxies 
and Compton-thick AGNs \citep[see][]{shi06,mushotzky93,wu09,georgantopoulos11}. In fact, we 
fitted the silicate feature at 9.7$\mu$m of the Compton-thick NGC\,3281 \citep{sales11},
using the {\sc clumpy} torus models \citep{nenkova02,nenkova08a,nenkova08b}, and found that 
the hydrogen column density derived from silicate profile is similar to that derived from 
X-Ray spectrum originally employed to classify this galaxy as Compton-thick source 
\citep[see also][]{shi06,mushotzky93,thompson09}.

Such result is further supported by the finding of \citet{shi06} which using observations 
of 9.7$\mu$m silicate features in 97 AGNs found that the strength of the silicate feature 
correlates with the H{\sc i} column density estimated from fitting the X-ray data, with 
high H{\sc i} columns corresponding to silicate absorption while low ones correspond to 
silicate emission. On the other hand, % it is worth mentioning that the literature results are controversial, 
\citet{thompson09}, for instance, suggested that even more informative than the 9.7$\mu$m 
feature alone the combination of it with the 18$\mu$m silicate feature reveals the geometry 
of the reprocessing dust around the AGNs, discriminating between smooth and clumpy distributions 
\citep[see also][]{sirocky08}, moreover, comparing 31 Sy~1 spectra obtained with IRS of 
21 higher luminosity QSOs, these authors conclude that the weak emission lines observed are 
a consequence of clumpy AGN surroundings. In addition, \citet{goulding12} studying the 20 nearest 
bona fide Compton-thick AGNs with hard X-ray measurements, shows that only about half of 
nearby Compton-thick AGNs have strong Si-absorption features and conclude that the dominant 
contribution to the observed MIR dust extinction observed is not solely related to the 
compact dusty obscuring structure surrounding the central engine but it can be originated from 
the host galaxy instead.

This paper is a part of a project that investigate the possibility that the presence of a silicate absorption 
feature at 9.7$\mu$m would be a signature of an heavily obscured AGN, we present here a study 
of the nuclear spectrum of Mrk\,3, a SB0 galaxy hosting an optically classified Seyfert~2 
(Sy~2) nucleus with a BLR detected in polarized light \citep{adam77,miller90,tran95,collins05}. 
\citet{capetti95} show that the narrow line region (NLR) of Mrk\,3 has a S-shaped morphology,
extended over nearly 2\arcsec, with a large number of resolved knots. They suggest that 
this morphology may be a consequence of the strong interaction between the NLR with the 
radio emission plasma \citep[see][]{capetti95,ruiz01,schmitt03}.

It is also known that Mrk\,3 has a complex X-ray spectrum with heavily-absorbed, and cold
reflection components, accompanied by a strong iron K$\alpha$ line at
$\sim$6.4\,keV \citep{awaki91,awaki08,cappi99,turner97,sako00}. \citet{awaki08} obtained an
intrinsic 2-10\,keV luminosity of $\sim1.6\,\times\,10^{43}$ erg s$^{-1}$, and suggested it was direct
emission from Mrk\,3. In addition, they found that there is a heavily absorbed dust/gas component of
N$_{H}\sim1.1\,\times\,10^{24}$\,cm$^{-2}$ obscuring the direct line of sight to the nucleus, 
and lead them to classify Mrk\,3 as a Compton-thick galaxy \citep[see also][]{awaki90,awaki91,iwasawa94,sako00}.

Nevertheless, the derived value of the hydrogen column density of Mrk\,3 is arguable, in the
light of studies developed by \citet{winter09}, who have shown that this target reveals a 
complex X-ray spectrum and a peculiar position in the color-color diagram of $F_{0.5-2keV}/F_{2-10keV}$ 
versus $F_{14-195keV}/F_{2-10keV}$, suggesting that its has high column density with a complex
changin-look \citep[see][for more details]{winter09}.

In this paper we present ground based, high spatial resolution, MIR spectra of the Compton-thick 
galaxy Mrk\,3. Such observation allowed the dust distribution to be studied in the central 
$\sim$200\,pc of this galaxy. As stated above the main goal is to investigate if the presence 
of a silicate absorption feature at 9.7$\mu$m of Mrk\,3 can be interpreted as a signature of an heavily 
obscured AGN caused by the dusty torus of the unified model. In addition, we briefly discuss 
the connection of dusty torus material with Compton-thick scattering material found in Sy~2 
Mrk\,3 and NGC\,3281. 

This paper is organized as follows: in Section~\ref{observation}, we briefly describe the 
observations and data reduction; in Section~\ref{results}, we discuss the results. Concluding 
remarks are given in Section~\ref{conclusions}.

\section{Observations and Data Reduction}\label{observation}

Michelle is a MIR ($7-26\mu$m) imager and spectrometer with a 320$\times$240 pixels Si:As 
IBC array. When configured as a long slit spectrometer it a plate scale of 0\farcs183/pixel
and a 21\farcs6 long slit. The low-resolution ($R\sim200$) long slit mode was used, with 
the 2-pixel wide (0\farcs366) slit, resulting in a dispersion of 0.024\,$\mu$m pixel$^{-1}$ 
and a spectral resolution of 0.08$\mu$m. The spectral coverage is $\sim$7$\mu$m centred at 
9.5$\mu$m. The total on-source integration time was 10 minutes.

\begin{figure*}
\centering
\includegraphics[scale=0.5]{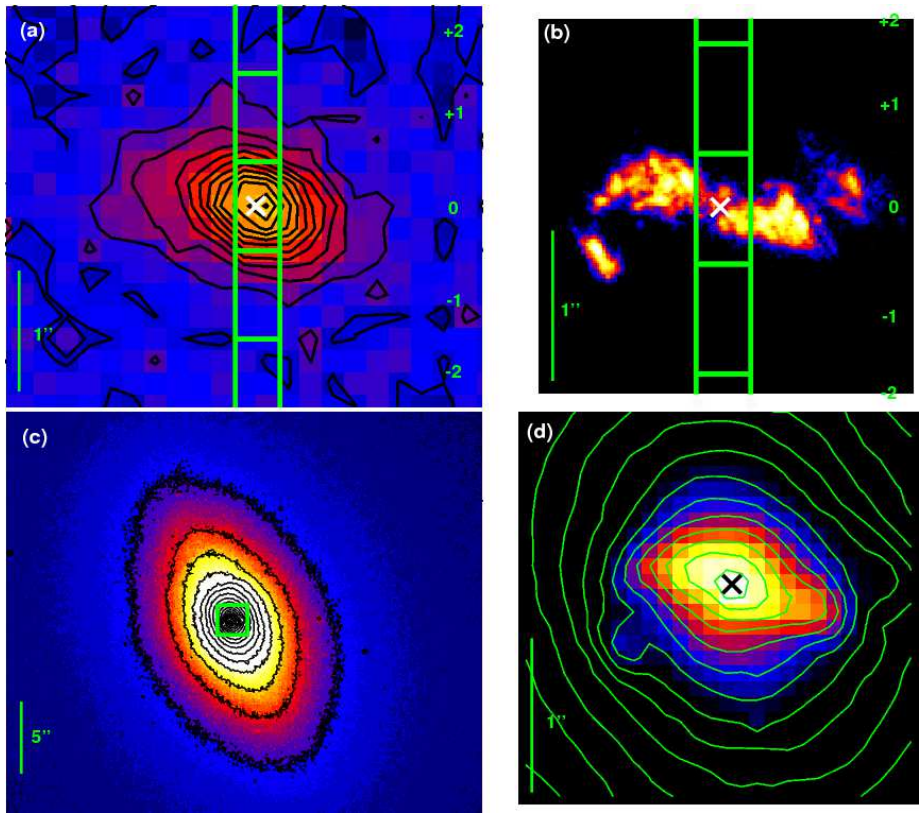}
\caption{(a) The long-slit position of Michelle overlays on the contoured Mrk\,3 acquisition 
image, which are linear and stepped at 5\% of the peak. (b) The [O{\sc\,iii]}\,$\lambda$5007\AA\,
image of Mrk\,3 which was observed using HST/FOC and taken from \citet{schmitt03b}. The
spectroscopic slit of Michelle are superposed. (c) The large scale optical continuum (F814W filter) 
obtained from the HST/WFPC2 data archive (Program 8645, PI: Windhorst, R.). The green box shows
the central 2\farcs5x2\farcs5 field of view and (d) its respectively amplification. The 
contours are logarithmically separated by a factor of 1.8. North is up and East is to the left. 
The position of the nucleus, measured in the HST/F814W continuum images, is plotted as a white and black cross.}
\label{mrk_fenda}
\end{figure*}

The data were obtained in queue mode at Gemini North, in 2010 February 05 UT, as part of 
program GN-2009B-Q-61 (PI: Pastoriza, M.). We obtained acquisition images with on-source time roughly 
5s through the $N$-band and Mrk\,3 was observed under clear conditions (photometric/cloudless), 
with precipitable waver vapour (PWV) in range of 2.3mm or less. The long-slit orientation,
%the observational conditions were photometric with the cloud cover,
%water vapour, and background being roughly 50\%, 20\% and 50\% respectively. The long-slit orientation,
as well as the extraction area of each spectrum are superposed upon the $N$-band acquisition 
and [O{\sc\,iii]}\,$\lambda$5007\AA\ \citep{schmitt03b} images of Mrk\,3 (Fig.~\ref{mrk_fenda}a,b).
We also present in Figure~\ref{mrk_fenda}c, the HST/WFPC2 archive optical continuum image 
(F814W filter, Program 8645, PI: Windhorst, R.). The green box overplotted on this panel 
represents the field of view of 2\farcs5x2\farcs5 and its amplification is shown in Fig.~\ref{mrk_fenda}d.

The image quality was measured from the $N$-band acquisition image of the telluric standard 
HD45866 and turns out to be $\sim$0\farcs7 (Fig.~\ref{profile}). In the same figure, we have also 
shown the spatial emission profile of Mrk\,3 from two directions, North to South (dash line) and East to 
West one (dash-dot line). Note that the spatial profile of the North to South (N-S) direction 
corresponds to our chosen position of the Michelle's long slit (see Fig.~\ref{mrk_fenda}).
Both emission profiles, the N-S and East to West (E-W) directions, were normalized using the 
average flux of the two central pixels (0\farcs366) of Mrk\,3.

A standard chop/nod technique was used to remove time-variable sky background, telescope 
thermal emission, and the effect of 1/\textit{f} noise from the array/electronics. The long-slit 
was oriented along P.A.\,=\,0\degr\, with a chop throw of 15\arcsec, oriented along P.A.\,=\,90\degr,
which includes only the signal of the guided beam position in the frame and avoid possible nod 
effects in the spatial direction. The same slit position/nod orientations were used for the 
telluric standards.

Data reduction employed the {\sc midir} and {\sc gnirs} sub-packages of Gemini IRAF
\footnote{IRAF is distributed by the National Optical Astronomy Observatory, which is 
operated by the Association of Universities for Research in Astronomy (AURA), Inc., under 
cooperative agreement with the National Science Foundation.} package. The final spectrum 
was extracted from the combined the chop- and nod-subtracted frames using the tasks {\sc tprepare} 
and {\sc mistack}. Wavelength calibration was obtained from the skylines in the raw frames. 
Removal of the telluric absorption lines from the galaxy spectrum, and the flux calibration 
were performed using the {\sc mstelluric} task, selecting HD\,45866 \citep{cohen99} as telluric
standard, observed immediately after the science target.

As seen in Fig.~\ref{profile}, Mrk\,3 has a luminosity profile similar to that of the
point like source meaning that the bright AGN emission could dominate the total emission,
however, there can be a contamination of an issuance arising from an extranuclear location. 
In the light of these thought, we performed two extraction from the 
2D-spectral image of Mrk~3 aiming estimate the contribution of the host galaxy around the AGN.

Therefore, first extraction was taken as the stellar FWHM (an apertures of four pixels 
or 0\farcs732 which correspond to $\sim$193\,pc for a distance of 55Mpc, using $H_{0}$ = 74 
km s$^{-1}$ Mpc$^{-1}$) and the other one is a spectrum of the whole emission available on 
the Michelle's long-slit ($\sim$1\,kpc). The host galaxy emission spectrum was derived by 
multiplying the central spectrum by a constant 1.314, further subtracting the result from the
spectrum of the largest possible aperture. This accounts for the infinite aperture correction of the FWHM extraction 
of Gaussian profile \citep[see][]{ruschel14}. We show in Fig.~\ref{mrk3} both the nuclear and host galaxy spectra.
The nuclear spectrum, unresolved point like source, was chosen to be centred at the peak 
of the continuum at $\approx11.3\mu$m which also coincides with the optical peak (see Fig.~\ref{mrk_fenda}).

The unresolved central spectrum of Mrk\,3 clearly reveals the silicate absorption feature at 
9.7$\mu$m as well as prominent forbidden lines of [Ar{\sc\,iii]}\,8.9\,$\mu$m, [S{\sc\,iv]}\,10.5\,$\mu$m 
and [Ne{\sc\,ii]}\,12.8\,$\mu$m, although the PAH emission is completely
absent (Fig.~\ref{mrk3}). The corresponding spectrum of the host 
galaxy neither provides emission lines nor silicate features, but this spectrum really shows 
low signal to noise to drive any conclusive analysis.

%From the 2D-spectral image of Mrk~3, we were able to extract 3 one dimensional spectra along the slit
%with apertures of four pixels or 0\farcs732 (which correspond to $\sim$193\,pc for a distance 
%of 55Mpc, using $H_{0}$ = 74 km s$^{-1}$ Mpc$^{-1}$) each one. The nuclear spectrum was 
%chosen to be centred at the peak of the continuum at $\approx11.3\mu$m which also coincides 
%with the optical peak (see Fig.~\ref{mrk_fenda}). The other four extractions were centred at 193\,pc, 
%and 386\,pc to the North (N), and South (S) directions (see Fig.~\ref{mrk3}).

\section{Results and Discussion}\label{results}

\subsection{Acquisition Image of Mrk\,3: Mid-Infrared Emission in the Narrow Line Region}\label{acquisition_image}

As it can be seen in Fig.~\ref{mrk_fenda}a, Mrk\,3 shows resolved prominent wings with 
extended emission in a roughly East to West direction centred on the unresolved nucleus. 
This emission lobe has a position angle (P.A.) of $\sim$70$^\circ$ and the easterly wing 
extended within $\sim$1.5\arcsec, while the westerly minor knot is elongated to $\sim$0.7\arcsec.
The emission in N-S direction is unresolved with a FWHM quite similar to that of the 
telluric star, $\approx$0\farcs7, which might suggest that the bulk of emission perpendicular 
to the wing direction is dominated by the central unresolved bright source implying that 
the putative torus can be the source of the central emission (see Section~\ref{sec:torus}).
On the other hand, we should note that the spatial resolution of the data is not sufficient to resolve 
the compact ($<0\farcs7$) point-like source on the N-S direction meaning that the torus emission 
could come from smaller region, pursuant to we need data with higher spatial resolution
to address this assumption.

%, meaning that the bulk of emission perpendicular to the 
%wing direction is dominated by the central unresolved bright source. It might imply that 
%the putative torus is effectively a point source within the central 0\farcs7 (see Section~\ref{sec:torus}).

\begin{figure}
\centering
\includegraphics[width=8cm]{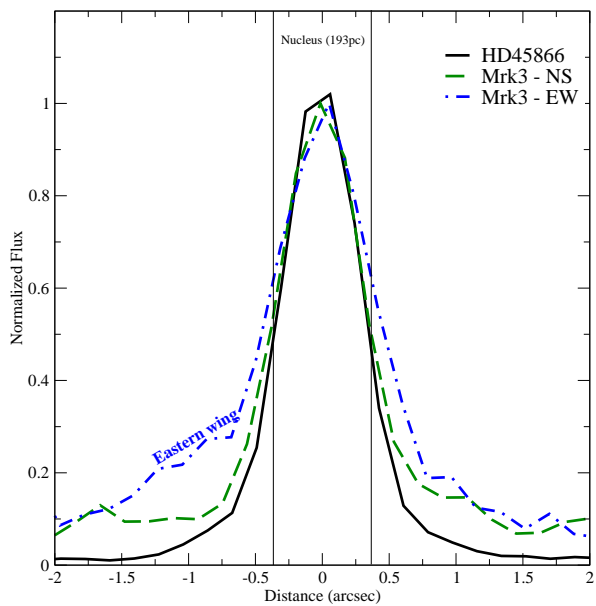}
\caption{Spatial emission profile of Mrk\,3 along to N-S (dash line), and E-West (dash-dot line) 
directions compared with that of the telluric standard star (HD\,45866, solid line).
The fluxes were normalised to the peak value. The positions of each spectral extractions are labelled.}
\label{profile}
\end{figure}

Mrk\,3 is a Sy 2 galaxy with a spectacular S-shaped 
NLR in [O{\sc\,iii]}\,$\lambda$5007\AA\, emission image (Fig.~\ref{mrk_fenda}). Its biconical 
region is extended toward $\sim$2\arcsec along the E-W direction and has been interpreted
as the result of a rapid expansion of a cocoon of hot gas heated by radio emission
\citep{capetti95,schmitt03}. Curiously our $N$-band acquisition image (Fig.~\ref{mrk_fenda}ab) 
mimics a cylindrical shell aligned with the ionization cone of [O{\sc\,iii]}\,$\lambda$5007\AA\, line.
The same assumption can be drawn from the collapsed emission profile (Fig.~\ref{profile})
measured in the central 0\farcs366 along to E-W direction. It is also notable that the Eastern 
wing is the brightest and the most extended emission, 1\arcsec, of Mrk\,3 at N-band wavelength.
Both, the spatial position and open angle of this emission feature, from N-band coincide with
enhanced east wing presents in the HST image of [O{\sc\,iii]}\,$\lambda$5007\AA\, emission line.

The $N$-band emission structure is also spatially enclosed within the extended emission present  
in the optical continuum image of the HST/WFPC2 F814W filter (see Fig.~\ref{mrk_fenda}cd). 
In order to highlight the extended NLR we also overploted contours on the optical continuum 
within the same field of view (2\farcs5 x 2\farcs5) of [O{\sc\,iii]}\,$\lambda$5007\AA\, image.
The spatial coincidence of the extended MIR emission with the observed size of the high 
excitation emission region might suggest that the dust present in the NLR of Mrk\,3 contributes 
to N-band emission. Indeed, the dust located within the ionization cone which maybe heated by the
central engine can contribute to the emission at MIR wavelength as it has been observed in 
several other AGNs \citep[e.g.][]{reunanen10,mason06,packham05,radomski02,radomski03}.

\begin{table}
\footnotesize
\renewcommand{\tabcolsep}{1.8mm}
\caption{N-band Flux Density of Mrk\,3\label{nflux}}
%\begin{minipage}[b]{1.0\linewidth}
\begin{tabular}{lccc}
\noalign{\smallskip}
\hline\hline
Radius  &	Flux Density &	Surface Brightness & Total Flux\\
(arcsec)&	(Jy) 	    &	(Jy arcsec$^{-2}$) & (percent)\\
\hline
0.366	&	0.21 &		0.51	& 36\%\\
0.732	&	0.44 &		0.26	& 73\%\\
1.098	&	0.55 &		0.15	& 91\%\\
1.464	&	0.59 &		0.09	& 97\%\\
1.83	&	0.60 &		0.06	& 100\%\\
\hline
\hline
\multicolumn{4}{l}{Errors in the flux calibration are accurate to be roughly 15\% }\\
\multicolumn{4}{l}{according to \citet{mason06,packham05}.}\\
%Position & Label &\multicolumn{2}{c}{\underline{~~~~~~~~[Ar{\sc\,iii]}$8.9\mu$m~~~~~~~~}} & \multicolumn{2}{c}{\underline{~~~~~~~~[S{\sc\,iv]}$10.5\mu$m~~~~~~~~}} & \multicolumn{2}{c}{\underline{~~~~~~~~[Ne{\sc\,ii]}$12.8\mu$m~~~~~~~}}& S$_{sil}$ & A$_{V}^{app}$ & $\tau_{9.7}$ & A$_{V}$ \\
%&&Flux&EW&Flux&EW&Flux&EW&&(mag)&&(mag)\\
%\noalign{\smallskip}
%\hline
%\noalign{\smallskip}
%193\,pc S & -1 & - & - & - & - & 0.001\,$\pm$\,0.0003 & 0.038 & -0.5\,$\pm$\,0.2 & 9.2\,$\pm$\,1.9 & 1.7\,$\pm$\,0.8 & 32\,$\pm$\,5\\
%Center     & 0 & 0.023\,$\pm$\,0.008 & 0.038 & 0.022\,$\pm$\,0.006 & 0.049 & 0.027\,$\pm$\,0.004 & 0.062 & -0.3\,$\pm$\,0.1 & 5.5\,$\pm$\,1.7 & 1.3\,$\pm$\,0.2& 24\,$\pm$\,6\\
%193\,pc N & 1 & 0.014\,$\pm$\,0.002 & 0.165 & 0.007\,$\pm$\,0.001 & 0.121 & 0.003\,$\pm$\,0.001 & 0.058 & -0.2\,$\pm$\,0.1 & 4.5\,$\pm$\,1.7 & 1.4\,$\pm$\,0.5& 26\,$\pm$\,2\\
\end{tabular}
%\end{minipage}
\end{table}

In Table~\ref{nflux} are listed the N-band flux density and surface brightness measured
with aperture upon different radius (see column 1). From these values, we can detected two
region of emitting dust. The brightest one arise from the central 0\farcs366 ($\sim 193$\,pc) 
region which contributed with 36\% of the total flux measured inside radius of 1\farcs83,
as well as with the bulk of the N-band emission (surface brightness 0.51 Jy arcsec$^{-2}$,
see column 3 in Tab.~\ref{nflux}). This result suggests that such component which dominates 
the unresolved point like source might emerge from the dusty torus predicted by unified model 
\citep[see also][]{packham05,radomski02,radomski03}. We call the attention that the used
spatial resolution does not resolve the torus emission, thus the Silicate emission comes 
from the central 200\,pc. But it is clear that outside this region the emission of Silicate
if it exists, is less than the detection limit of our Gemini/Michelle data.

The acquisition N-band image also shows a second component that is a extended diffuse emission with lower surface brightness 
extending up to 800\,pc from the nucleus. This extended emission, which coincides with the 
NLR seen in the HST image of [O{\sc\,iii]}\,$\lambda$5007\AA\, (Fig.~\ref{mrk_fenda}), is 
therefore probably due to the dust emission located in the NLR and heated by the central 
source of Mrk\,3.

\subsection{Silicate absorption and Emission Lines Measurements}\label{decomposition}

\begin{figure*}
\centering
\includegraphics[scale=0.6,angle=-90]{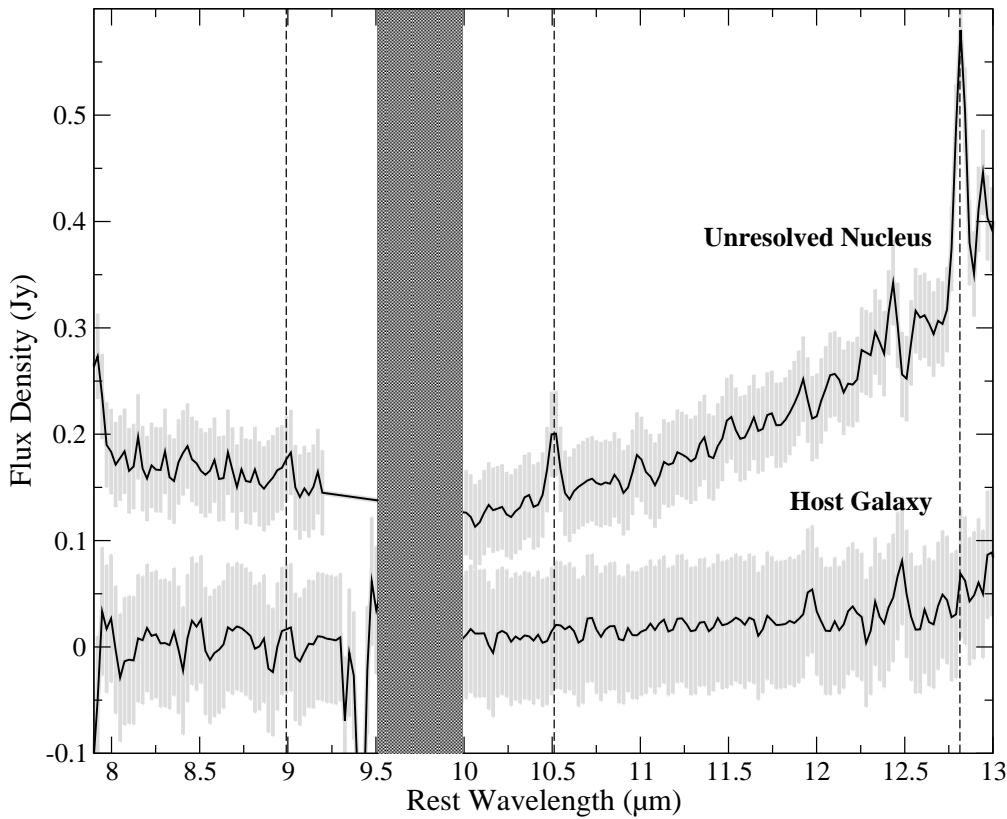}
\caption{{\bf Spectra of the unresolved nucleus (0\farcs732 or 193\,pc) as well as the host
galaxy of Mrk\,3. %observed with Gemini/Michelle. %Two offset extraction centred on the 193\,pc to the North and South directions are also shown (top panel, black lines). 
%The integrated low resolution spectrum of Mrk\,3 observed with Spitzer/IRS, extracted from 
%an aperture of radius 7\farcs2, is presented in the bottom panel as blue line. 
Dotted lines show the [Ar{\sc\,iii]}\,8.9\,$\mu$m, [S{\sc\,iv]}\,10.5\,$\mu$m and [Ne{\sc\,ii]}\,12.8\,$\mu$m 
ionic lines. Grey shared area is the root-mean-square deviation (RMSD) and the hatched area represent 
the O$_{3}$ telluric band.}}
\label{mrk3}
\end{figure*}

According to \citet{hao07}, the presence of silicate absorption feature at 9.7$\mu$m and no 
evidence of PAH emission bands could be interpreted as a galaxy with a heavily obscured 
active nucleus \citep[see also][]{shi06,mushotzky93,thompson09,sales10,sales11}. This picture 
seems to be the case of our nuclear spectrum of Mrk\,3 (Fig.~\ref{mrk3}) which presents a 
clear absorption feature at 9.7$\mu$m, as well as strong ionic lines at [Ar{\sc\,iii]}\,8.9\,$\mu$m, 
[S{\sc\,iv]}\,10.5\,$\mu$m and [Ne{\sc\,ii]}\,12.8\,$\mu$m, but no PAH have been observed.

%The similar characteristics appear on the spectrum observed using the Spitzer/IRS telescope 
%\citep[see Fig.~2 of][]{weedman05}. However, the integrated low resolution spectrum observed 
%by Spitzer/IRS, which was extracted from a circular aperture of radius 7\farcs ($\approx 2$\,kpc), 
%clearly shows an increase in flux towards longer wavelengths ($>\,9.5\mu$m) suggesting an 
%extended extra-nuclear component of dust emission emerging from the AGN host galaxy. 

On the other hand, the high resolution extended spectrum of Mrk\,3 obtained with Gemini/Michelle 
does not reveal such increase in flux, setting not only a lower limit for the galactocentric 
distance of the dust emission observed by Spitzer/IRS, but also corroborate with the claim that the AGN torus of
Mrk\,3 is constrained within the central $\sim 200$\,pc extraction.

%set reveal two scenario: (i) there is no significant 
%emission in the neighbourhood ($>193$\,pc) of the unresolved nucleus or (ii) no detectable 
%signal at resolution element of our data.

%This effect may emerge colder (T) dust component produce a 

%Note that the spectra of the external positions (193\,pc N and S) show similar features as 
%those in the unresolved nucleus, although much less prominent (see Fig.~\ref{mrk3}).

%ionic lines
%The central extraction that corresponds to nuclear point like source of Mrk\,3 
%The spectra clearly show a silicate absorption 
%feature at 9.7$\mu$m, as well as strong [Ar{\sc\,iii]}\,8.9\,$\mu$m, [S{\sc\,iv]}\,10.5\,$\mu$m 
%and [Ne{\sc\,ii]}\,12.8\,$\mu$m ionic lines in the nuclear extraction.
%Both spectra at 193\,pc N and S show the same features as those in the unresolved nucleus,
%although less prominent (see Fig.~\ref{mrk3}). In addition, none of the positions show 
%polycyclic aromatic hydrocarbon (PAH) bands, in agreement with the Spitzer spectrum \citep{weedman05}. 
%According to \citet{hao07} the presence of silicate absorption but no evidence of PAH emission
%bands, could be interpret as a heavily obscured active nucleus \citep[see also][]{shi06,mushotzky93,thompson09,sales10,sales11}.

Following the procedure described by \citet{spoon07}, we adjusted a liner continuum centred 
at 8 and 12.5$\mu$m to derive the intrinsic, unobscured, AGN flux at 9.7$\mu$m. From that, 
the apparent strength ($S_{sil}$) of the silicate absorption has been calculated as:

\begin{equation}
S_{sil} = ln\,\frac{f_{obs}(9.7\,\mu m)}{f_{cont}(9.7\,\mu m)}, 
\end{equation}
where $f_{obs}$ and $f_{cont}$ are the observed and the intrinsic AGN unobscured fluxes, 
respectively. Subsequently, we estimated the silicate optical depth, $\tau_{9.7}$, by the 
correlation of $\tau_{9.7}=-S_{sil}$ \citep{nenkova08b}. The apparent optical extinction 
attributed to silicate dust can be inferred from A$_{V}^{app}$ = $\tau_{9.7}\,\times$\,18.5\,$\pm$\,2\,mag 
\citep{draine03}. The resulting values are listed in Tab.~\ref{lines}.

We estimated a value of $S_{sil}$ = $-0.52\,\pm\,0.21$ to the unresolved nucleus of Mrk\,3.
This value is consistent with the average value observed in Sy\,2 galaxies \citep[$\langle\,S_{sil}\rangle=-0.61$, see Fig.~2 of][]{hao07}.
From $S_{sil}$ we derived an optical dust extinction of A$_{V}^{app}$ = 9.6\,$\pm$\,0.51\,mag.

In order to disentangle the contribution from each component in the spectral energy distribution 
(SED) of Mrk\,3 as a function of the distance to the nucleus, we used the {\sc pahfit}
\footnote{Source code and documentation for {\sc pahfit} are available at http://tir.astro.utoledo.edu/jdsmith/research/pahfit.php} 
IDL routines \citep{smith07}. This code decomposes the spectrum as continuum emission from dust 
and starlight, emission lines, individual and blended PAH emission bands, and also
assumes that the light is attenuated by extinction due to silicate grains. The code uses 
the dust opacity law of \citet{kemper04}, and the infrared extinction is considered as a 
power law plus silicate features peaking at 9.7\,$\mu$m.

\begin{figure}
 \begin{centering}
%  \begin{tabular}{c}
%   \subfigure[ ]{\includegraphics[clip=true,width=8cm]{Mrk3_l1_4pix.eps}}\\
\includegraphics[clip=true,width=8cm]{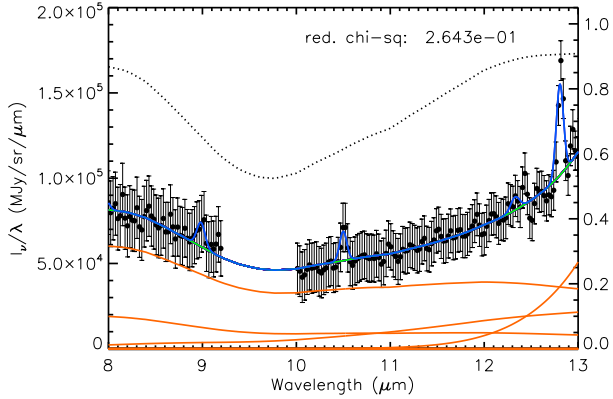}\\
%   \subfigure[ ]{\includegraphics[clip=true,width=8cm]{Mrk3_r1_4pix.eps}}\\
%  \end{tabular}
%   \par
 \end{centering}\vspace{-0.3cm}
\caption{Decomposition of the centre spectrum of Mrk\,3 using {\sc pahfit} code. The best fit model is represented 
by the blue line. The dotted black line indicates the mixed extinction components, while 
the orange lines represent the individual thermal continuum components.}
\label{spectra_pahfit}
\end{figure}

The {\sc pahfit} code requires information of the uncertainties, and here those were assumed 
to be the root-mean-square deviation (RMSD) estimated from the whole spectral range of Michelle's 
N-band presented as a grey shaded regions in Fig.~\ref{mrk3}. Since the spectra do not show PAH 
%10\% of the observed flux, as expected for Michelle observations 
%\citep[see][]{radomski02,mason06,sales11,sales12}. Since the spectra do not show PAH 
emission, this component was not included in the fitting \citep[see][]{sales11}. 
The resulting 
spectral decomposition of the nuclear extraction of Mrk\,3 is shown in Fig.~\ref{spectra_pahfit}, 
and the fluxes derived to the emission lines are given in Tab.~\ref{lines}.

%It is clear from Fig.~\ref{spectra_pahfit} that the silicate absorption strength (dotted 
%black line) varies along the slit, which reflect in different values for $\tau_{9.7}$. 
The unresolved nucleus shows $\tau_{9.7}$ = 1.46$\pm$0.4. %, while it reaches $\tau_{9.7}$ = 1.11$\pm$0.8
%and $\tau_{9.7}$ = 1.94$\pm$0.7 to the 193\,pc N and S extractions, respectively. 
Inspecting the measured dust extinction (A$_{V}$) values obtained using {\sc pahfit}, one 
can see that it is larger than those inferred from the S$_{sil}$. Such difference arises 
because the S$_{sil}$ are obtained from the peak of the silicate feature at 9.7$\mu$m while the 
{\sc pahfit} code takes into account the whole silicate profile ($8 - 14\mu$m). %Thus, the S$_{sil}$ 
%indicator should be used only as a lower limit for A$_{V}$. In addition, 
Note that the optical depth values are similar to those found in other Seyfert galaxies \citep{gallimore10}.
%We did not find a gradient of temperature through the spectral extractions of Mrk\,3
%and their thermal dust continuum has been represented by a black-bodies with T = 350 and 300\,K.
%However, we should keep in mind on discussion presented here is that the neighbourhood 
%extraction (193\,pc N and S) may contain a significant amount of the light scattered from the central 
%source once that we have chosen it to be the FWHM of the Gaussian profile which means only 
%roughly 76\% of the whole flux of the target.

%The thermal dust continuum components that best fits the unresolved nucleus and 193\,pc N 
%is a black-body with T = 800\,K, while the 193\,pc S spectrum is best represented by a 
%black-body of T = 150\,K. This colder region is more heavily obscured (A$_{V}$ = 32\,$\pm$\,5\,mag) 
%than the nucleus and the N regions. However, we do not find evidence of the additional optical 
%reddening in the [O\,{\sc iii}]$\lambda$5007\AA\ maps at this position \citep{schmitt03b}.

\begin{table}
\large 
\renewcommand{\tabcolsep}{2.5mm}
\caption{Derived Parameters from MIR Feature of Mrk\,3.\label{lines}}
%\begin{minipage}[b]{1.0\linewidth}
\centering
\begin{tabular}{lcc}
\noalign{\smallskip}
\hline\hline
Features & $\lambda$ ($\mu$m) & Centre \\ %& 193\,pc S & 193\,pc N\\
\hline
F\,[Ar{\sc\,iii]} & 8.9 & 7.82\,$\pm$\,1.9 \\ %& - & 4.08\,$\pm$\,3.2\\
EW\,[Ar{\sc\,iii]} & 8.9 & 0.039 \\ %& - & 0.049\\
%FWHM\,[Ar{\sc\,iii]}$8.9\mu$m &&&\\
F\,[S{\sc\,iv]} & 10.5 & 7.60\,$\pm$\,1.8 \\ %& 0.42\,$\pm$\,0.25 & 1.75\,$\pm$\,0.75\\
EW\,[S{\sc\,iv]} & 10.5 & 0.051 \\ %& 0.061 & 0.082\\
%FWHM\,[S{\sc\,iv]}$10.5\mu$m &&&\\
F\,[Ne{\sc\,ii]} & 12.8 & 13.2\,$\pm$\,2.5 \\ %& 0.67\,$\pm$\,0.24 & 1.06\,$\pm$\,0.61\\
EW\,[Ne{\sc\,ii]} & 12.8 & 0.055 \\ %& 0.036 & 0.030\\
%FWHM\,[Ne{\sc\,ii]}$12.8\mu$m &&&\\
S$_{sil}$ & 9.7 & -0.52\,$\pm$\,0.21 \\ %& -0.85\,$\pm$\,0.5  & -0.61\,$\pm$\,0.5\\
A$_{V}^{app}$ & - & 9.6\,$\pm$\,0.51 \\ %& 15.8\,$\pm$\,2 & 11.34\,$\pm$\,1\\
$\tau$\,$^{a}$ & 9.7 & 1.46\,$\pm$\,0.4 \\ %& 1.94\,$\pm$\,0.7 & 1.11\,$\pm$\,0.8\\
A$_{V}$\,$^{a}$ & - & 27\,$\pm$\,0.4 \\ %& 36\,$\pm$\,0.5 & 20.5\,$\pm$\,0.9\\
\hline
\hline
\multicolumn{3}{l}{Fluxes are in units of 10$^{-4}$ W\,m$^{-2}$\,sr$^{-1}$.}\\
\multicolumn{3}{l}{EWs are in $\mu$m.}\\
\multicolumn{3}{l}{$^{a}$ Values derived from {\sc pahfit} code.}\\
%Position & Label &\multicolumn{2}{c}{\underline{~~~~~~~~[Ar{\sc\,iii]}$8.9\mu$m~~~~~~~~}} & \multicolumn{2}{c}{\underline{~~~~~~~~[S{\sc\,iv]}$10.5\mu$m~~~~~~~~}} & \multicolumn{2}{c}{\underline{~~~~~~~~[Ne{\sc\,ii]}$12.8\mu$m~~~~~~~}}& S$_{sil}$ & A$_{V}^{app}$ & $\tau_{9.7}$ & A$_{V}$ \\
%&&Flux&EW&Flux&EW&Flux&EW&&(mag)&&(mag)\\
%\noalign{\smallskip}
%\hline
%\noalign{\smallskip}
%193\,pc S & -1 & - & - & - & - & 0.001\,$\pm$\,0.0003 & 0.038 & -0.5\,$\pm$\,0.2 & 9.2\,$\pm$\,1.9 & 1.7\,$\pm$\,0.8 & 32\,$\pm$\,5\\
%Center     & 0 & 0.023\,$\pm$\,0.008 & 0.038 & 0.022\,$\pm$\,0.006 & 0.049 & 0.027\,$\pm$\,0.004 & 0.062 & -0.3\,$\pm$\,0.1 & 5.5\,$\pm$\,1.7 & 1.3\,$\pm$\,0.2& 24\,$\pm$\,6\\
%193\,pc N & 1 & 0.014\,$\pm$\,0.002 & 0.165 & 0.007\,$\pm$\,0.001 & 0.121 & 0.003\,$\pm$\,0.001 & 0.058 & -0.2\,$\pm$\,0.1 & 4.5\,$\pm$\,1.7 & 1.4\,$\pm$\,0.5& 26\,$\pm$\,2\\
\end{tabular}
%\end{minipage}
\end{table}

\section{Dusty Torus Constraints from the Central Region of Mrk\,3}\label{sec:torus}

In the unified model for AGN, the presence of a nuclear toroidal structure composed by 
dust/gas-rich matter, mainly silicate and graphite, is postulated to attenuate the nuclear 
emission at UV/optical wavelengths and re-emit it in the MIR, which leave unmistakable 
signatures in the observed SEDs. While the sublimation of the graphite grains creates IR 
emission at $\lambda\ge1\mu$m,  the $\sim$9.7$\mu$m feature observed in emission/absorption 
is attributed to silicate grains \citep[e.g.][]{barvainis87,pier92,granato94,siebenmorgen05,fritz06,ardila06,riffel06,riffel09}.  
Some authors consider the torus as a continuous density distribution \citep[e.g.][]{pier92,granato97,siebenmorgen04,fritz06}, 
but it has been postulated that, for dust grains to survive in the torus environment, they 
should be shielded within clumpy structure \citep{krolik88}, which simultaneously provides 
a natural attenuation of the silicate feature \citep[e.g.][]{nenkova02,nenkova08a,nenkova08b,honig06}.

As can be noted on Fig.~\ref{profile} the emission profile within two central pixels (0\farcs366, 
the same width of Michelle long-slit) is quite similar to that of the telluric standard 
HD\,45866 showing that the dust emission in the NLR is not resolved at this spatial resolution.
However, we may infer worth information modelling the emission from the unresolved nucleus.
With this purpose, after masking the emission lines and the telluric band region (Fig.\ref{mrk3}) 
using a simple interpolation we subsequently compared the unresolved spectrum to {\sc clumpy}\footnote{The 
models are available at http://www.pa.uky.edu/clumpy/} theoretical SEDs \citep{nenkova02,nenkova08a,nenkova08b}. 
These models assume a complex clumpy dust distributions of the toroidal geometry of AGN unified 
scheme which is constrained by six parameters (\textit{i}) number of dusty clouds along to 
the torus equatorial ray, \textit{$N_{0}$}; (\textit{ii}) the visual optical depth of each 
clump, $\tau_{V}$; (\textit{iii}) radial extension of the clumpy distribution, $Y=R_0/R_d$, 
where $R_0$ and $R_d$ are the outer and inner torus radii, respectively; (\textit{iv}) the 
radial distribution of clouds as described by a power law $\propto r^{-q}$; (\textit{v}) the
torus angular width follows a Gaussian angular distribution described by a half-width $\sigma$; 
(\textit{vi}) the observer's viewing angle $i$.

In order to constrain the six torus parameters from {\sc clumpy} model we used the {\sc BayesCLUMPY} 
tool developed by \citet{asensioramos09}. Its statistic methodology is based on Bayesian 
inference framework and performs a Markov Chain Monte Carlo (MCMC) search over the parameter 
grid. This approach has become commonplace for SED fitting because of its economy and efficiency 
over an exhaustive search of large parameter spaces.

\begin{figure}
\centering
\includegraphics[width=8cm]{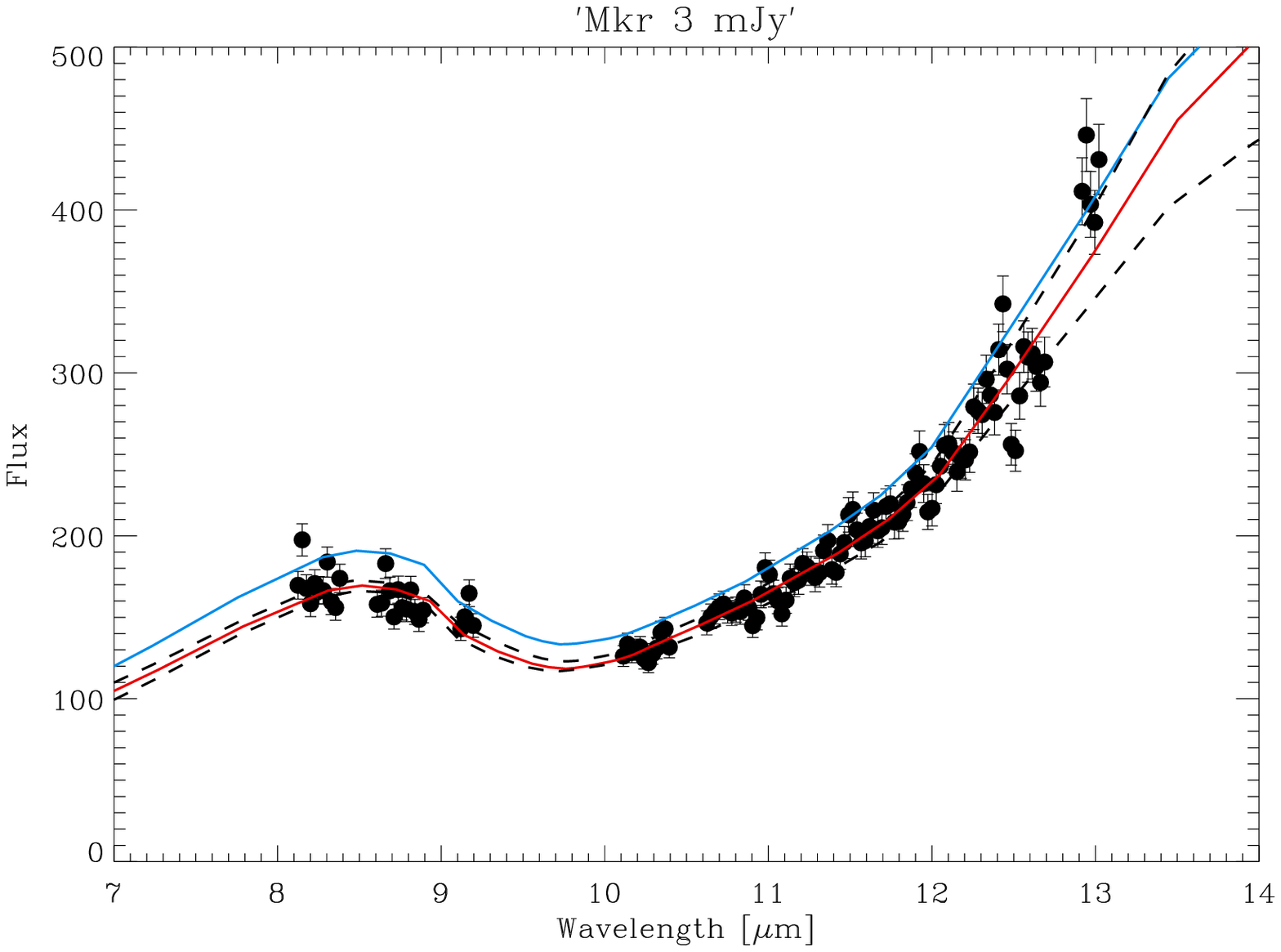}
\includegraphics[width=8cm]{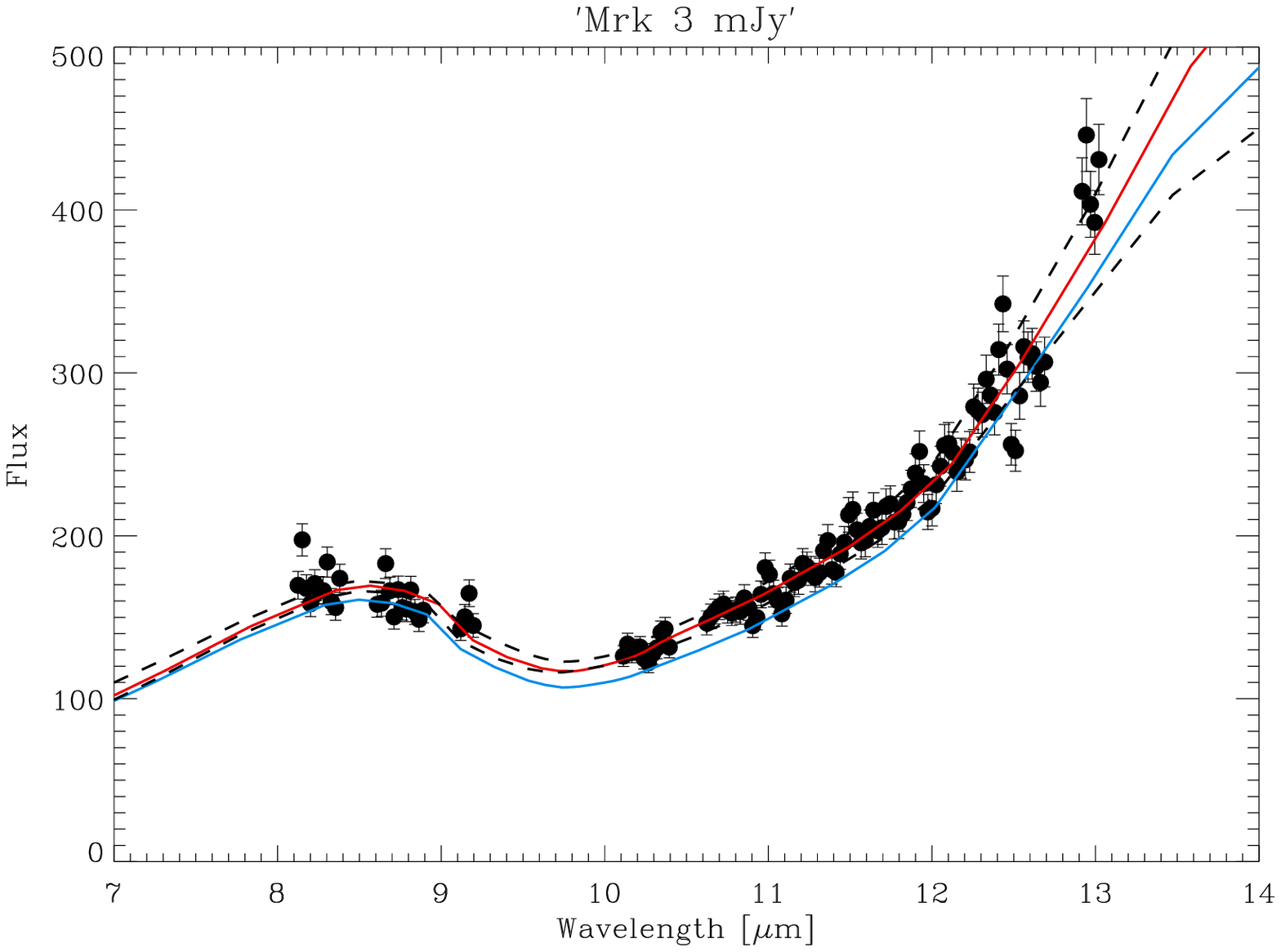}
\caption{$N$-band fitting of Mrk\,3 using a clumpy torus model. The blue line
represents the best fit model taken from the values that maximizes the probability distributions (MAP) of
the parameters, while red line describes median value of each parameter. The dashed lines
show the upper and lower limits of 68\% confidence interval for each parameter around the median.}
%nuclear spectrum (solid black line), after removal 
%of the underlying galaxy contribution, and the best fit torus model (red line) as described 
%in the text, normalised at 9.0$\mu$m. The grey dotted lines correspond to the locus of the 
%10\% best-fitting SEDs.}
\label{nenkova}
\end{figure}

\begin{figure}
\centering
\includegraphics[scale=0.43]{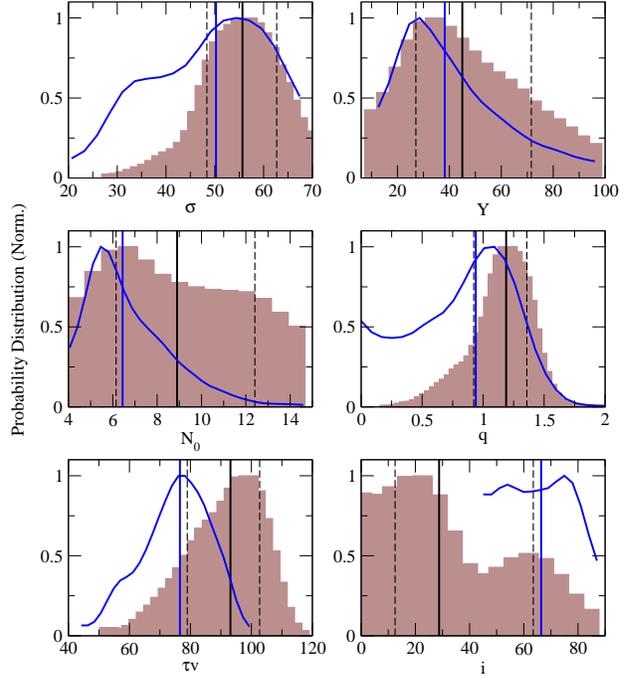}
\caption{Probability distributions of the fitted parameters using Nenkova's clumpy torus model 
of Mrk\,3 with the observer angle of view, $i$, taken as free parameter (histograms) and constrained
to be between $45^{\circ}\le\,i\,\le\,90^{\circ}$ (blue curve).
The solid lines, black and blue, represent the best models that maximizes their probability distributions (MAP).
The dashed lines indicate the 68\% confidence level for each parameter around the MAP.
The marginal posterior distributions have been normalised to unity.}
\label{nenkova_hist}
\end{figure}

{\sc BayesCLUMPY} takes into account the first 13 eigenvectors of the principal component 
analysis, which is a very good representation of the whole $\sim 10^6$ model grid of {\sc clumpy}
SED database \citep[see][]{asensioramos09}, to describe the observed spectra. The marginalized 
probability distributions are assessed as histograms of the Markov Chain for each parameter 
derived from the automatic marginalization properties of the MCMC technique. We perform two 
running sets (Fig.~\ref{nenkova}), the first one used all six torus parameters as free while 
the second run limited the observer's line of sight angle to be $i\ge$45$^{\circ}$ \citep[see][for details]{alonsoherrero11}.
The stability of the solution has been confirmed by consecutive runs of the algorithm and
the results are shown in Fig.~\ref{nenkova_hist} and Tab.~\ref{tab:fit_model_mrk3}.

\begin{table}
\centering
%\scriptsize
%\small
\footnotesize
\renewcommand{\tabcolsep}{1mm}
\caption{{\sc clumpy} Torus Parameters of Mrk\,3}
\label{tab:fit_model_mrk3}
\begin{tabular}{lccc}
\hline\hline
\noalign{\smallskip}
\noalign{\smallskip}
Parameter & Best Fit$^{a}$ & \multicolumn{2}{c}{\underline{68\% confidence interval}}\\
&&Upper & Lower \\
\hline
Angular width ($\sigma$) & 56$^{\circ}$ & 48$^{\circ}$ & 63\\
Radial thickness ($Y$) & 45 & 27 & 71 \\
Number of clouds ($N_0$) & 9 & 6 & 12 \\
Index of radial density ($q$) &	1.18 & 0.92 & 1.35 \\
Observer angle ($i$) & 29$^{\circ}$ & 12$^{\circ}$ & 63$^{\circ}$ \\
Cloud optical depth ($\tau_V$) & 93\,mag & 79\,mag & 103\,mag \\
%\multicolumn{4}{l}{Bolometric Luminosity, L$_{bol}\approx$ 1.82$\times10^{44}$erg/s}\\
%\multicolumn{4}{l}{Inner, R$_d\sim0.17$pc, and outer, R$_0\sim8$pc, radii}\\
%\multicolumn{4}{l}{Hydrogen column density $N_{obs}\sim2.74\times10^{23}$}\\
\hline
\multicolumn{4}{c}{Observer angle constrained as edge-on view ($45^{\circ}\le\,i\,\le\,90^{\circ}$)}\\
\hline
Angular width ($\sigma$) & 50$^{\circ}$ & 35$^{\circ}$ & 61$^{\circ}$\\
Radial thickness ($Y$) & 38 & 25 & 63 \\
Number of clouds ($N_0$) & 6 & 5 & 9 \\
Index of radial density ($q$) &	0.94 & 0.39 & 1.25 \\
Observer angle ($i$) & 66$^{\circ}$ & 53$^{\circ}$ & 70$^{\circ}$ \\
Cloud optical depth ($\tau_V$) & 76\,mag & 64\,mag & 87\,mag \\
%\multicolumn{4}{l}{Bolometric Luminosity, L$_{bol}\approx$ 2.17$\times10^{44}$erg/s}\\
%\multicolumn{4}{l}{Inner, R$_d\sim0.19$\,pc, and outer, R$_0\sim7$\,pc, radii}\\
%\multicolumn{4}{l}{Hydrogen column density $N_{obs}\sim4.76\times10^{23}$}\\
\hline
\hline
\multicolumn{4}{l}{$^{a}$The value parameters that maximizes their probability (MAP)}\\
\multicolumn{4}{l}{distributions}\\
%Total number of solution & 	 & 3	& 20& 55& 95\\
\noalign{\smallskip}
\end{tabular}
%\caption*{The value parameters that maximizes their probability (MAP) distributions}
%\footnotetext[1]{The value parameters that maximizes their probability (MAP) distributions}
\end{table}

Following our approach the {\sc clumpy} model was able to reproduce very well the $N$-band spectrum 
of Mrk\,3 using either observer view angle, $i$, as free parameter (first run) or fixed in Sy~2 edge-on view 
(second run, Fig.~\ref{nenkova} and Tab.~\ref{tab:fit_model_mrk3}) predicted by the unified model of AGN \citep{antonucci93}.
These modelling claim a presence of an embedded AGN surrounding by a dusty toroidal structure 
constraining the radial thickness to $R_0/R_d$ = 45$^{+26}_{-18}$, $i$ as free, or 38$^{+25}_{-13}$ to edge-on view.
%The parameters obtained above suggest that the nucleus of Mrk\,3 hosts a dusty toroidal 
%structure, with a radial thickness of $R_0/R_d$ = 80. 
The first runs turned out that the Mrk~3's torus has 9$^{+3}_{-3}$ dusty clouds in its equatorial 
radius with the individual clouds presenting an optical depth of $\tau_{V}$ = 93$^{+10}_{-14}$\,mag, 
while the best fit on the second run was reached with 6$^{+3}_{-1}$ clouds, with $\tau_{V}$ = 
76$^{+11}_{-12}$\,mag instead.
%its equatorial radius with the individual clouds 
%presenting an optical depth of $\tau_{V}$ = 93\,mag and 76 respectively.
%The density distribution of the clouds follows a power-law in the form r$^{-1.18}$, 
%and the toroidal structure is constrained in angular width by a Gaussian distribution of 
%$\sigma$ = 56$^{\circ}$ for the first run, while r$^{-0.9}$ and $\sigma$ = 50$^{\circ}$
%have been obtained on the second run.%gotten to constrained Sy~2 view instead. 

Once we have forced the observer's line of sight to vary between 45 and 90$^{\circ}$, following
the assumption that Sy~2 is always viewed close to the torus's plane, the $i$ parameter turns roughly 66$^{+4}_{-13}$\,degree,
while using $i$ as free parameter we reached $i = 29^{+34}_{-17}$\,degree. 
%The model also indicates that we are looking through the viewer angle of $i$ = 66$^{\circ}$
%with respect to torus geometry, once that we constrained the observer line of sight to vary
%between 45 and 90$^{\circ}$ in order to carry up the Sy 2 angles of Mrk\,3. 
%In order to estimate the hydrogen column density in the observer's view we determine the 
The hydrogen column density along the observer's view is estimated by adopting a hydrogen 
column density of N$_{H}\approx10^{23}$\,cm$^{-2}$ (upper limit value) for each cloud \citep{nenkova02,nenkova08a,nenkova08b}
and the number of cloud over the observer line of sight as follows
\begin{equation}
N_{obs} = N_0\,\,\,exp \left[ - \frac{(90-i)^2}{\sigma^2}\right],
\end{equation}
leading to an upper limit value of $N_{H}^{obs} = 2.7^{+7.2}_{-2.2}\,\times\,10^{23}$\,cm$^{-2}$ 
and 4.8$^{+3.3}_{-3.1}\,\times\,10^{23}$\,cm$^{-2}$ using results from the first and second 
runs (see Fig.~\ref{nenkova_hist} and Tab.~\ref{tab:parameters_clumpy}).

%if we adopting a hydrogen 
%column density of N$_{H}\approx10^{23}$ for each cloud \citep{nenkova02,nenkova08a,nenkova08b}. 
%and the hydrogen column density in the observer's line of sight is N$_{H}\,\approx\,1.4\,\times\,10^{24}$\,cm$^{-2}$ 
%\citep[see][]{nenkova08b,sales11}. 

Using the equation \citep{nenkova08b}:
\begin{equation}
R_d = 0.4\times\left(\frac{1500K}{T_{sub}}\right)^{2.6}\left(\frac{L_{bol}}{10^{45}erg/s}\right)^{0.5}\,pc
\end{equation}
and adopting a dust sublimation temperature of $T_{sub}=$1500\,K we obtained the torus 
size as roughly 8$^{+4}_{-3.4}$\,pc and 7$^{+5}_{-2.2}$\,pc from run 1 and 2, respectively.

\begin{table}
\centering
\scriptsize
%\small
%\footnotesize
\renewcommand{\tabcolsep}{1mm}
\caption{Parameters derived from {\sc clumpy} constraints}
\label{tab:parameters_clumpy}
\begin{tabular}{lcc}
\hline\hline
\noalign{\smallskip}
\noalign{\smallskip}
Parameter & $0^{\circ}\le\,i\,\le\,90^{\circ}$ & $45^{\circ}\le\,i\,\le\,90^{\circ}$\\
\hline
Bolometric Luminosity, L$_{bol}$ (erg/s) & 1.82$\times10^{44}$ &  2.17$\times10^{44}$ \\
X-ray luminosity, L$_{X-ray}$ (erg/s) & 9.1$\times10^{42}$ & 1.35$\times10^{43}$ \\
Hydrogen column density, N$_{H}^{obs}$ (cm$^{-2}$) & 2.7$^{+7.2}_{-2.2}\,\times\,10^{23}$ & 4.8$^{+3.3}_{-3.1}\,\times10^{23}$\\
Direct view of AGN, P$_{esc}$ (\%) & 6 & 0.85 \\
Inner radius of torus, R$_d$ (pc) & 0.17 & 0.19 \\
Outer radius of torus, R$_0$ (pc) & 7.68$^{+4}_{-3.4}$ & 7.08$^{+5}_{-2.2}$ \\
\hline\hline
\noalign{\smallskip}
\end{tabular}
%\caption*{The value parameters that maximizes their probability (MAP) distributions}
%\footnotetext[1]{The value parameters that maximizes their probability (MAP) distributions}
\end{table}

The scale amplitude of the {\sc clumpy} model best fit to the N-band observed spectrum turns out 
the bolometric flux of the AGN. From these, we derived a bolometric luminosity of 
L$_{bol}$ = 1.82\,$\,\times\,10^{44}\,$erg s$^{-1}$ for the first run and 
L$_{bol}$ = 2.17\,$\,\times\,10^{44}\,$erg s$^{-1}$ using the second one. 
In order to compare with X-ray 
observations we converted it, using a correction factor of 20 as proposed by \citet{elvis94}, 
to a total X-ray luminosity which reaches a value of L$_{X-ray}$$\approx\,9.1\,\times\,10^{42}\,$erg s$^{-1}$
and L$_{X-ray}$$\approx\,1.35\,\times\,10^{43}\,$erg s$^{-1}$.% \citet{awaki08} inferred to 

%{\bf 
%It is worth mention that the intrinsic degeneracy of the fitting {\sc clumpy} torus
%models leads us to reproduce the same observed spectrum using different model sets. This 
%issue is well notable once that we are using MIR narrow band observed on the ground-based.
%However, $N$-band spectra cover a rich wavelength range in which we can study the most relevant 
%feature, silicate at 9.7$\mu$m, linked to the predicted torus of the AGN unified model.
%However discussing degeneracies between the relevant parameters is well outside the scope 
%of this work.
%}

\subsection{Comparison with Previous Observations}

%Mrk\,3 has been extensively studied and shows a bright AGN surrounded
%by a torus with edge-on view \citep[see][]{collins05,schmitt03,awaki08,pounds05,crenshaw10}. 
%It also shows ionization emission in the optical characterized by a Z shape that has been decomposed
%into a NLR and extended NLR (ENLR) by \citet{crenshaw10}. These analysis inhibited 
%the geometries of both NLR and ENLR as biconical with maximum half-opening angle of $\theta=51^{\circ}$
%and $\theta=28^{\circ}$ respectively. These authors have also pointed out that Mrk\,3 displays its
%inner dust/gas disk inclined around $I\sim64^{\circ}$. 

We compare the parameters of the torus derived from the run 1 and 2 with ones obtained from 
X-ray observed by \citet{awaki08,awaki90}. The results obtained with the second run, which 
fixed the edge-on torus view, of Mrk\,3 is favoured (see Tab.~\ref{tab:parameters_clumpy}) 
by the physical properties derived from X-ray observation. Indeed, 
the X-ray luminosity inferred from our modelling turns out a similar value to that derived 
from the X-ray data (see Tab.~\ref{tab:parameters_clumpy}). Using X-ray high quality wide-band 
(0.4-70\,keV) spectrum of Mrk\,3 observed with the Suzaku telescope \citet{awaki08} were able 
to derive an intrinsic X-ray luminosity of L$_{2-10keV}$ = 1.6\,$\times10^{43}\,$erg s$^{-1}$
for this galaxy, concluding that it is quite completely attenuated by Compton-thick cold 
matter with $N_H\sim\,1.1\times10^{24}$cm$^{-2}$. However, clumpy torus model give us a 
hydrogen column density, $N_{H}^{obs}\sim4.8^{+3.3}_{-3.1}\,\times10^{23}$, about twice smaller than 
those obtained from \citet[$N_H\sim\,0.7-1.1\times10^{24}$cm$^{-2}$,][]{awaki08,awaki90}.
They also inferred the existence of a Compton shoulder to the iron K$\alpha$ line, interpreted 
as the presence of a dusty torus. They also suggested that the iron emitting matter is located 
at $>$\,1\,pc from the hidden nuclear engine. On the other hand, the size of the scattering 
region, due to the ionized gas, extends roughly over 50\,pc \citep[see also][]{sako00,pounds05,serlemitsos07,awaki90}.

%In addition, as it has been discussed previously \citep[e.g.][]{elitzur08,ramos11,alonsoherrero11,nenkova02,mor09} 
%the clumpy distribution of the Sy~1 and Sy~2 types of AGNs will not depend on the observer's 
%viewing but of the probability that the observer actually seeing the hidden active engine. Hence,

Assuming that the probability of a photon escapes from the central source toward of viewing angle
$i$, in a clumpy environment composed by optically thick clouds ($\tau_{V}>1$), is determined 
by $P\,\simeq\,exp(-N_{obs})$, which $N_{obs}$ is the number of cloud over the observer's line of sight.
The first and second runs yield a $P\,\sim6\%$ and $P\,\sim0.85\%$ respectively. These results
suggest that our run with Sy~2 view recovers a scenario which the amount of ionizing photons from the AGN 
are absorbed by the torus cloud environment as predict by unified models and it might explain
the Sy~2 optical features observed in this galaxy.
However, it is difficult to set tightly constraints from our modelling once even not the high-spatial 
resolution of the Gemini/Michelle $N$-band spectrum could spatially resolve the
point like source of Mrk\,3 nor allows the contribution from dust in the host 
galaxy be estimated \citep[see][]{goulding12}.% or cover a wide spectral range of Mrk\,3.

\subsection{Compton-thick Torus Model Geometry of Seyfert 2: Mrk\,3 and NGC\,3281}

In the clumpy dust torus model for AGN obscuration discussed above, the observed strength 
of the silicate dust features is expected to be dependent on the specific geometry and 
optical depth of the torus \citep{nenkova02,fritz06,granato94}. In addition, the dust would
be co-spatial with cold neutral gas, which absorbs the X-ray wavelength light from the AGN 
\citep[see][]{shi06,mushotzky93,wu09}. In particular, \citet{shi06} found that the strength 
of the silicate feature correlates with the hydrogen column density, suggesting that the 
material surrounding the central engine is responsible for obscuring both the X-ray and the 
silicate emission. Therefore, \citet{georgantopoulos11} postulates that the presence of a 
strong MIR silicate feature in absorption is a good method to find the most heavily obscured 
AGN, also called Compton-thick sources.

Following that line, \citet{sales11} represented the torus model geometry of Compton-thick Sy~2 
NGC\,3281 using the {\sc clumpy} models, finding evidence that the matter responsible for 
the silicate absorption at 9.7\,$\mu$m could also be responsible for the X-ray scattering, 
with a column density of N$_{H}\sim1.2\,\times\,10^{24}$\,cm$^{-2}$. We also 
constrained parameters of the CLUMPY model to the torus geometry of Mrk\,3 
(see Section \ref{sec:torus}), which might indicate a similar correlation between the dusty 
material associated with the silicate feature and the X-ray scattering region in this galaxy. 
However, we should note that this conclusion may be subject of some concerns as it has been 
explored by \citet{goulding12} and need to be further addressed.
%In order to highlight their different environments, we present in Fig.~\ref{fig:torus_ngc3281_mrk3} 
%a scheme representing the characteristics of the NGC\,3281 and Mrk\,3 tori. 
The physical parameters of each torus are listed in Tab.~\ref{tab:torus_ngc3281_mrk3}.

From our analysis, we can conclude that the torus model geometry derived to NGC\,3281 and Mrk\,3 
are quite similar, which have only differences in the visual optical depth of clump's clouds 
as well as the total number of clouds in the torus equatorial radius suggesting that the dust 
torus in NGC\,3281 is more compact than Mrk\,3 (Tab.~\ref{tab:torus_ngc3281_mrk3}).

%in NGC\,3281 shows a large opening angle ($\sigma\,=\,70^{\circ}$), 
%and is rather compact ($Y\,=\,20$), Mrk\,3 hosts a quite extended ($Y\,=\,80$) and thin 
%($\sigma\,=\,15^{\circ}$) structure. We also see differences in the radial density profile, 
%outer radii, as well as the total number of clouds in the torus equatorial radius suggesting 
%that the dust torus in NGC\,3281 is more compact than Mrk\,3. As expected, the 
%viewing angle in both galaxies is compatible with a Sy~2 line of sight. 

If we take A$^{obs}_V = 1.086\,N_0\,\tau_V\,e^{-\left[\frac{(i - 90)^2}{\sigma^2}\right]}$ 
as an estimate of the optical extinction along the line of sight, we obtain A$^{obs}_V$\,=\,506\,mag 
for NGC\,3281, and A$^{obs}_V$\,=\,393\,mag for Mrk\,3. The difference in the optical 
obscuration generated by the two tori may explain the depth of the silicate profiles as the
spectrum of NGC\,3281 shows a deeper silicate feature than that of Mrk\,3 \citep[see Fig. 2 in][]{sales11}. 
It is important to highlight that these comparison between the torus model geometries of Mrk\,3
and NGC\,3281 is not straightforward because the fitting methodology of the clumpy models used 
by \citet{sales11} is different from that used here.

Furthermore, we can estimate the density of silicate matter as
\begin{equation}
\rho_{9.7\mu m} = \tau_{9.7\mu m}/\kappa L, 
\end{equation} 
where $\tau_{9.7\mu m}$ was obtained from the {\sc pahfit} analysis, $\kappa$ is the mass 
absorption coefficient of emitter material, and $L$ is emitting region length. Assuming 
$\kappa\,=\,315$\,m$^2$\,kg$^{-1}$, which corresponds to MgSiO$_3$ amorphous silicate 
\citep[see Tab.~5 in][]{whittet03}, and $L$ as the length of uncontaminated nuclear extraction 
(193pc), we obtain $\rho_{9.7\mu m}=7.1\,\times\,10^{-21}$\,kg\,m$^{-3}$ and
7.8$\,\times\,10^{-22}$\,kg\,m$^{-3}$ for NGC\,3281 and Mrk\,3 respectively. This 
difference in the silicate density values is in agreement with the trend found to the optical 
extinction inferred using the SEDs of the {\sc clumpy} models.

Note that the modelling  of the silicate feature at 9.7$\mu$m using \citet{nenkova02} clump 
torus also allow us to compare the hydrogen column density with the values derived from X-ray 
observations. Both NGC\,3281, (N$_{H}\,\approx\,1.5\,\times\,10^{24}\,cm^{-2}$) and Mrk\,3 
(N$_{H}\,\approx\,4.8^{+3.3}_{-3.1}\,\times\,10^{23}\,cm^{-2}$) have been classified as Compton-thick 
sources \citep[see also][]{vignali02,awaki08,sales11}. Therefore, the results from 
\citet{sales11} and the new results presented here likely provide further evidence that the 
silicate-rich dust is indeed associated with the AGN torus and possibly also responsible 
for the absorption observed at X-ray wavelengths. However, it is necessary higher spatial 
resolution from the data and further accuracy in its modelling.

Since Mrk 3 has a roughly face-on view \citep[$b/a=0.89$][]{goulding12}, 
we interpreted that the bulk of the N-band emission comes from
a region less than the unresolved central $\sim$200\,pc. In addition, we are
assuming that the dominant emission is related to the putative torus
of the AGN unified model. However, despite our claim, the high spatial
resolution (0\farcs366) of Gemini/Michelle is not enough to
isolate the Mrk 3 torus emission from its inner 200 pc. Hence, we have
to keep in mind that our nuclear spectrum may have a contribution of
the silicate dust located the inner 200 pc of the Mrk\,3 bulge as
suggested for other galaxies \citep[see][for review]{deo2007,deo2009,goulding2009,mullaney2011}.

Moreover, \citet{shi06} also %state that such correction, more specifically the scatter in the relation, 
suggest that the structure of the material surrounding the central black hole requires a multi-phase medium composed
by circumnuclear disk geometry plus a middle disk with a diffuse component and embedded denser
clouds as well as the clumpy outer disk. Hence, we need to keep in mind, that the direct relation
of the dust-to-gas ratio, as we have been used here, is causing concern, but this issue is 
not part of the scope of this paper and should be investigated more carefully in the future.

%\begin{figure}
%\centering
%\includegraphics[scale=1.6,angle=0]{torus_comp.eps}
%\caption{Cartoon comparing the characteristics of NGC\,3281 (top) and Mrk\,3 (bottom) tori 
%as derived in \citet{sales11} and this work. R$_d$ and R$_0$ are the inner and outer radii 
%respectively.}\label{fig:torus_ngc3281_mrk3}
%\end{figure}

\begin{table*}
\centering
\footnotesize
\renewcommand{\tabcolsep}{0.3mm}
\caption{Resulting Parameter for the Torus Models of NGC\,3281 and Mrk\,3}
\label{tab:torus_ngc3281_mrk3}
\begin{tabular}{lcc}
\hline\hline
Parameter & NGC\,3281 & Mrk\,3\\
\hline
Torus angular distribution width ($\sigma$)			& 70$^{\circ}$	&50$^{\circ}$  \\
Radial torus thickness ($Y$) 	    				& 20 		&38 \\
Clumps along to torus equatorial radius ($N_0$) 			& 10		&6\\
Power-law index of radial density ($q$) 			& 1.0		&0.94\\
Observer viewing angle ($i$)	 				& 60$^{\circ}$	&66$^{\circ}$\\
Clump visual optical depth ($\tau_V$)	 			& 40\,mag 	&76\,mag\\
Torus outer radii ($R_0$)					& 11\,pc	&7\,pc \\
Observer visual extinction (A$^{obs}_V$)			& 506\,mag	&393\,mag \\
MgSiO$_3$ amorphous silicate density ($\rho_{9.7\mu m}$)		& {~~~~~~~~7.1$\times10^{-21}$kg\,m$^{-3}$~~~~~~~~} & 7.8$\times10^{-22}$\,kg\,m$^{-3}$\\
Observer column density	($N_H$)					& {~~~~~~~~1.5$\times10^{24}$cm$^{-2}$~~~~~~~~}   & 4.76$\times10^{23}$cm$^{-2}$ \\
Nuclear extraction size 					& 65\,pc        & 193\,pc\\
\hline
\end{tabular}
\end{table*}

\section{Summary and Conclusions}\label{conclusions}

In this work we present a study using high spatial resolution ($\sim$193\,pc) 
spectra of the N-band wavelength (8--13$\mu$m) of the well known Compton-thick 
galaxy Mrk~3 in order to investigate the correlation between the Compton-thick 
material seen at X-ray wavelength and the silicate grain signature at 9.7$\mu$m. 
We also compare the results find here to Mrk~3 with those of Compton-thick galaxy 
NGC~3281, where the silicate absorption properties could be linked to Compton-thick 
material from X-ray spectra. Our main conclusions are:

\begin{enumerate}
  \item No polycyclic aromatic hydrocarbons (PAHs) emissions were detected in the Mrk~3 
  spectra. However, strong [Ar{\sc\,iii]}\,8.9\,$\mu$m, [S{\sc\,iv]}\,10.5\,$\mu$m and 
  [Ne{\sc\,ii]}\,12.8\,$\mu$m ionic emission-lines as well as a silicate absorption feature 
  at 9.7$\mu$m have been detected at the nuclear spectrum.% The same features, but much less 
  %intense, are observed in the North and South extractions centred 193\,pc from the nucleus. 
  
  \item By analysis of the N-band image of Mrk\,3 we are able to detected two emitting region,
  which the brightest one is dominated by the unresolved central source that might emerge from
  the dusty torus of the unified model. However, we should note that the spatial resolution
  of our Gemini/Michelle spectrum could not actually resolve the nuclear torus emission.
  The second component is an extended MIR emission from 
  the circumnuclear region of Mrk\,3. This diffuse dust emission shows up as a wings towards 
  E-W direction mimicking the same S-shaped of the Narrow Line Region as has been seeing 
  in the optical image of [O{\sc\,iii]}\,$\lambda$5007\AA.

%  \item The nuclear optical extinction, inferred from silicate feature at 9.7$\mu$m, is 
%  $A_V$=27$\pm$0.4\,mag. We have found larger value of the optical extinction in the 
%  South ($A_V$=36$\pm$0.5mag) extraction while smaller one to the 193\,pc North ($A_V$=20.5$\pm$0.9\,mag).

\item The nuclear spectrum was compared with $\sim10^6$ SEDs of {\sc clumpy} torus models, 
the result suggests that the nuclear region of the Mrk~3 hosts a dusty toroidal structure 
with an angular cloud distribution of $\sigma = 50^{+11}_{-15}$ degree, observer's view angle $i = 
66^{+4}_{-13}$ degree, and an outer radius of R$_{0}\sim$7$^{+5}_{-2.2}$\,pc. The hydrogen column density along 
the line of sight, derived from Nenkova's torus models, is N$_H\,=\,4.8^{+3.3}_{-3.1}\,\times\,10^{23}$\,cm$^{-2}$. 
The torus models also provide an estimate for the X-ray luminosity (L$_{X-ray}$ $\approx\,1.35\,\times\,10^{43}\,$erg s$^{-1}$) 
of the AGN in Mrk\,3 and this value is comparable to that derived from observed X-ray spectra, 
L = 6.2\,$\times10^{43}\,$erg s$^{-1}$.

\item By comparing the torus properties of Mrk~3 and NGC~3281 Compton-thick Sy~2 galaxies 
it turns out similar torus model geometries. This result perhaps indicates further evidence 
that the silicate dust is associated with the torus predicted by the unified model of AGN, 
and could also be responsible for the absorption observed at X-ray wavelengths of those galaxies 
classified as Compton-thick sources. However, it is necessary better spatial resolution in
order to address this assumption.

\end{enumerate}

\section*{Acknowledgements}
\label{acknowledgements}

MP would like to acknowledge the support from CNPq (grant 308985/2009-5). R.R. acknowledges 
funding from FAPERGs (ARD 11/1758-5) and CNPq (grant 304796/2011-5). Based on observations 
obtained at the Gemini Observatory, which is operated by the Association of Universities 
for Research in Astronomy, Inc., under a cooperative agreement with the NSF on behalf of the 
Gemini partnership: the National Science Foundation (United States), the Science and Technology 
Facilities Council (United Kingdom), the National Research Council (Canada), CONICYT (Chile), 
the Australian Research Council (Australia), Minist\'{e}rio da Ci\^{e}ncia e Tecnologia 
(Brazil), and Ministerio de Ciencia, Tecnolog\'{i}a e Innovaci\'{o}n Productiva (Argentina).

{}

\end{document}